\documentclass[12pt]{iopart}

\usepackage{graphicx}
\usepackage{epsfig}
\usepackage{harvard}
\usepackage{iopams}
\usepackage{color}

\usepackage{lineno}

\newcommand{\text}{\rm }

\begin{document}

\title[The Micromegas detector of the CAST experiment]{The Micromegas detector of the CAST experiment}

\author{ P~Abbon$^1$, S~Andriamonje$^1$, S~Aune$^1$,
  T~Dafni$^{2,7}$\footnote{Present address: DAPNIA, Centre d'\'Etudes
    Nucl\'eaires de Saclay (CEA-Saclay), Gif sur Yvette, France},
  M~Davenport$^3$, E~Delagnes$^1$, R~de Oliveira$^3$,G~ Fanourakis$^4$,
  E~Ferrer Ribas$^1$, J~Franz$^5$, T~Geralis$^4$, M~Gros$^1$,
  Y~Giomataris$^1$, I G~Irastorza$^1$, K~Kousouris$^4$, J~Morales$^6$,
  T~Papaevangelou$^3$, J~Ruz$^6$, K~Zachariadou$^4$, K~Zioutas$^{3,8}$}

\address{$^1$ DAPNIA, Centre d'\'Etudes Nucl\'eaires de Saclay (CEA-Saclay), Gif sur Yvette, France} \ead{eferrer@dapnia.cea.fr}
\address{$^2$ Technische Universit\"at Darmstadt, IKP, Schlossgartenstrasse~9, D-64289 Darmstadt, Germany}
\address{$^3$ European Organization for Nuclear Research (CERN), CH-1211 Gen\`eve 23, Switzerland}
\address{$^4$ National Center for Scientific Research "Demokritos", Athens, Greece}
\address{$^5$ Universit\"at Freiburg, Physikalisches Institut, Herrman-Herder-Strasse 3, D-79104 Freiburg, Germany}
\address{$^6$ Instituto de F\'{\i}sica Nuclear y Altas Energ\'{\i}as, Universidad de Zaragoza, Zaragoza, Spain}
\address{$^7$ Gesellschaft f\"ur Schwerionenforschung, GSI-Darmstadt, Plasmaphysik, Planckstr. 1, D-64291 Darmstadt}
\address{$^8$ University of Patras, Patras, Greece}

\begin{abstract}
  A low background Micromegas detector has been operating in the CAST
  experiment at CERN for the search of solar axions during the first phase
  of the experiment (2002-2004). The detector, made out of low
  radioactivity materials, operated efficiently and achieved a very low
  level of background rejection ($5\times 10^{-5}$ counts
  keV$^{-1}$cm$^{-2}$s$^{-1}$) without shielding.
\end{abstract}


\submitto{\NJP}

\maketitle

\section{Introduction}
The CAST experiment \cite{andriamonje:07b,zioutas:05a} uses three different
types of detectors to detect the X-rays originated from the conversion of
the axions inside a magnet:~a time projection chamber
\citeaffixed{autiero:06a}{TPC,}, an X-ray telescope \cite{kuster:06a}, and
a Micromegas detector. The Micromegas detector of CAST is a gaseous
detector optimized for the detection of low energy ($1$--$10\,\text{keV}$)
X-ray photons. It is based on the micropattern detector technology of
MICROMEGAS (MICROMEsh GAseous Structure) developed in the mid 90's
\cite{giomataris:96a,giomataris:98a,charpak:02a}. \citeasnoun{collar:00a}
first suggested the advantages of using the Micromegas for such
low-threshold, low-background measurements as required by the CAST
experiment. These advantages include sensitivity at the keV and sub keV
energy region where very good energy resolution can be achieved, excellent
spatial resolution, one dimensional or X-Y readout capability, stability,
construction simplicity and low cost. In addition, the proper choice of
construction materials would lead to a detector appropriate for low
background measurements.

The CAST Micromegas group designed and constructed a low background
detector, the very first made with an X-Y readout structure, optimized for
the efficient detection of the $1$--$10\,\text{keV}$ photons.  Several
detectors have been developed during the course of the CAST running, each
new one with increasingly improved characteristics replacing the older
module during shutdown and maintenance periods.  The detector is mounted on
one of the two west superconducting magnet apertures looking for 'sunrise'
axions converted into X-ray photons that will enter the detector active
volume perpendicularly to the X-Y strip plane.

\section{Detector description}
\begin{figure}
  \centerline{\includegraphics[width=7cm]{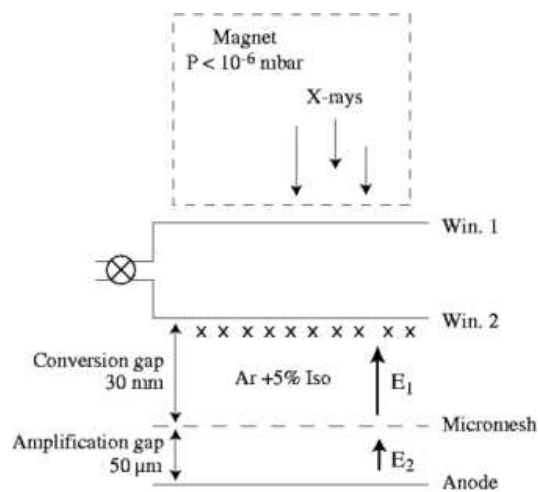}}
  \caption{The Micromegas CAST detector.} \label{fig:mmCAST}
\end{figure}

The principle of operation of the Micromegas detector designed for the CAST
experiment is sketched in \fref{fig:mmCAST}. A photon, after traversing a
vacuum buffer space, enters the conversion-drift region, filled with a
mixture of Argon-Isobutane ($95\%$--$5\%$), where it generates a
photoelectron via the photoelectric effect. The photoelectron travels a
short distance during which it creates ionization electrons. The electrons
drift in a field of about 250~V$/$cm, until they reach and funnel through
the micromesh and into the amplification region where a strong field of
about $40\,\text{kV}\,\text{cm}^{-1}$ causes an avalanche. The resulting
electron cluster is collected on the X-Y strips of the anode plane. The
maximum achievable gain is about $10^5$, but for CAST gains of $5 \times
10^3$ up to $10^4$ are sufficient to achieve the required threshold (around
$1\,\text{keV}$).
The main sources of background are cosmic rays and natural radioactivity.
Special care has been taken in the materials used for the construction of
this Micromegas detector in order to reduce the natural radioactivity.
Other developments have been necessary in order to optimize this detector
given the aim and the environment of the experiment. A description of the
most important elements specific to the CAST Micromegas detector are given
below.

\subsection{Mechanical structure}
\begin{figure}
  \centerline{\includegraphics[width=0.6\textwidth]{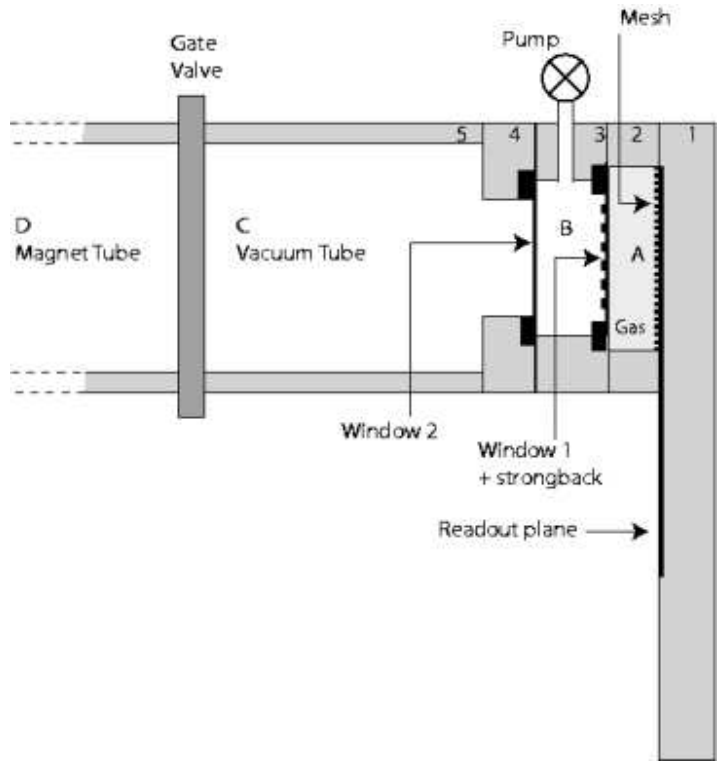}}
  \caption{The mechanical design of the detector. The drift electrode
    is attached on disk~3. The micromesh holds on disk~2 and disk~4 is used
    to hold an extra vacuum window. The drawing is not made to scale.}
  \label{fig:mmMech}
\end{figure}

The detector frame consists of Plexiglas cylinders held together via
plastic bolts. The drift and multiplication electrodes are attached on
these cylinders. \Fref{fig:mmMech} shows the mechanical structure of the
detector. The conversion region can be $2.5$ or $3\,\text{cm}$ thick and is
formed between a $4\,\mu\text{m}$ thick aluminized polypropylene window,
glued on stainless steel or aluminum strongback, capable of holding vacuum
at the magnet side, and the micromesh plane. The window of the conversion
region also serves as the cathode for the drift field. The amplification
region is only $50(100)\,\mu\text{m}$ thick and is formed between the
micromesh plane and the charge collection plane with the help of pillars
spaced $1\,\text{mm}$ apart. The micromesh is made of $4\,\mu\text{m}$
thick Copper and is fabricated at CERN \cite{delbart:02a}.

\subsection{Differential pumping}
The detector is fastened to one of the magnet bores with the help of an
aluminum tube and a flange. A gate valve separates the magnet volume from
the tube volume. In order to couple a gaseous detector with a vacuum
environment, keeping the maximum transparency to X-ray photons and a
minimum vacuum leak, the solution of two windows with a differential
pumping was adopted. The two windows are made out of a thin film of
$4\,\mu\text{m}$ polypropylene. The first window, that undergoes a pressure
difference of $1\,\text{bar}$, is glued on a strongback with a 94.6\%
transparency.  The two windows delimit 3 zones that can be seen in
\fref{fig:mmMech}.  Zone~A is the gaseous detector at a pressure of
$1\,\text{bar}$.  Zone~B is the vacuum gap at a pressure of
$5\times10^{-4}\,\text{mbar}$ obtained with the pumping group.  Zone~C is
the vacuum tube at a pressure of $5\times10^{-7}\,\text{mbar}$ in the
magnet.  The leak of the first window is proportional to the differential
pressure between zone~A and~B, i.e., $1\,\text{bar}$. This differential
pressure imposes the use of a strongback. The leak for this window due to
its porosity, tested with zone~A full of helium, is
$4\times10^{-5}\,\text{mbar}\,\text{l}\,\text{s}^{-1}$.  As the
differential pressure between zones~B and~C is only
$5\times$10$^{-4}\,\text{bar}$, a strongback is not needed.  The net leak
for this window when zone~A is full of helium, has been measured to be
$3\times10^{-9}\,\text{mbar}\,\text{l}\,\text{s}^{-1}$. The leak on the
first window has been evacuated by the pump. The pump system used for this
application is made of a small dry turbo pump (magnetic bearing) and of a
dry primary pump. The convolution of the transmission of the two windows
together with the conversion efficiency of photons in the detector gas
(Argon with 5\% Isobutane) over the energy spectrum of solar axions between
$2$ and $10\,\text{keV}$ results in an combined efficiency of 85\%. For
sub-kev sensitivity a more efficient gas, like Xenon, could be used as well
as thinner polypropylene windows.

\subsection{Charge collection in two dimensions}
\begin{figure}
  \begin{minipage}{0.49\textwidth}
    \centerline{\includegraphics[width=0.9\textwidth]{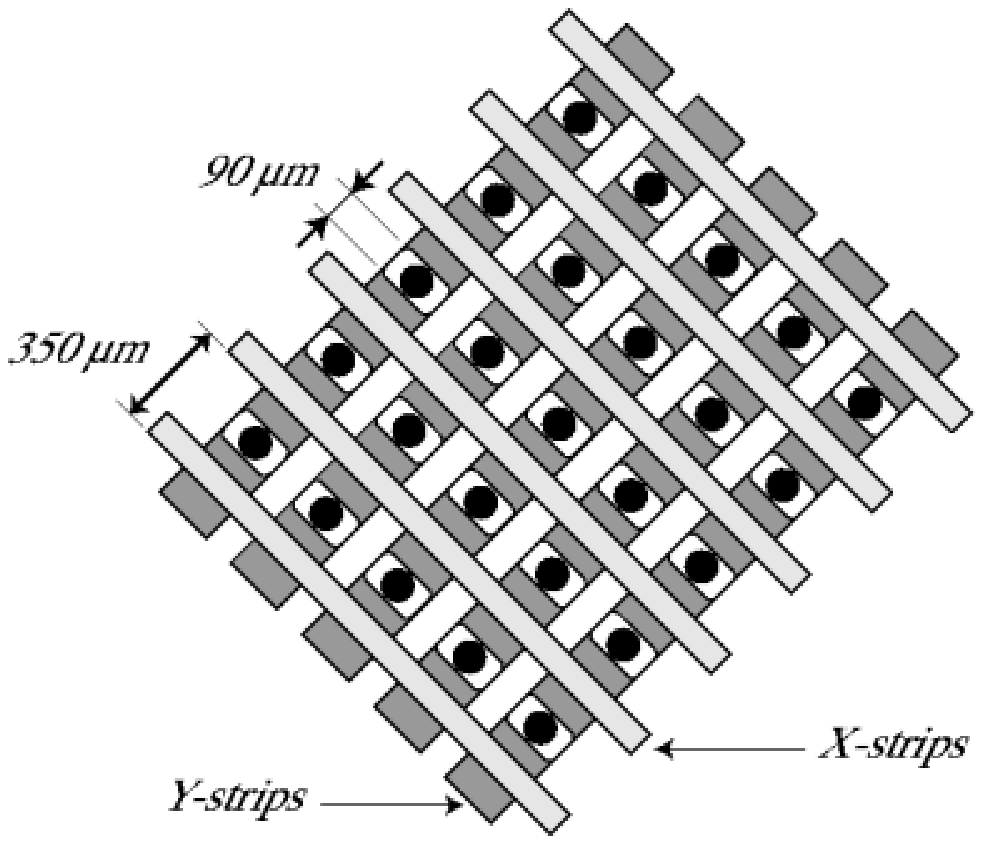}}
  \end{minipage}
  \hfill
  \begin{minipage}{0.49\textwidth}
    \centerline{\includegraphics[width=0.9\textwidth]{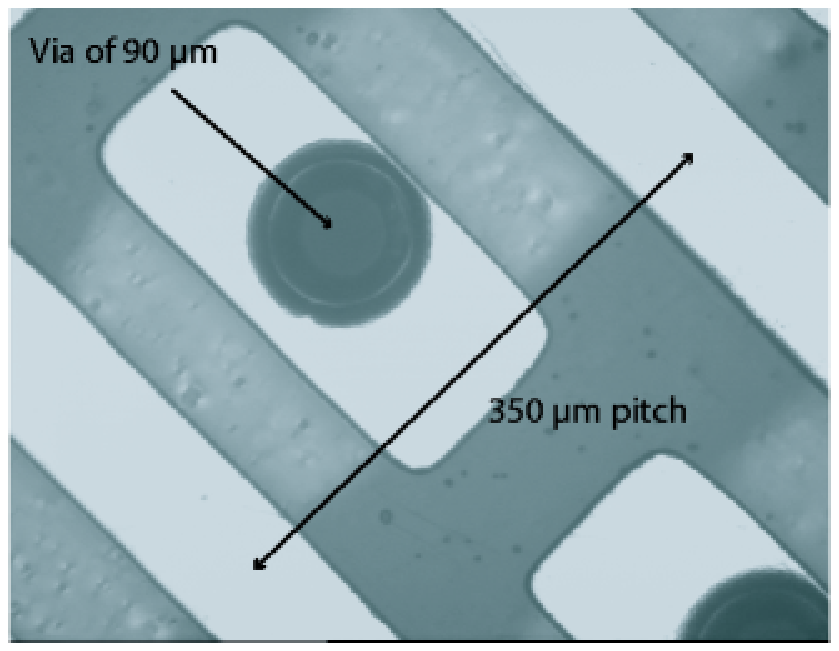}}
  \end{minipage}
  \caption{The X-Y strip charge collection structure. The strip pitch
    is $350\,\mu\text{m}$. The X srips are those in light grey and the
    Y-strips, in the underneath layer, in dark grey. The metalized holes of
    90~$\mu$m, allow the surface charge collection for the Y-strips.}
  \label{fig:2dreadout}
\end{figure}

The charge collection strips comprise an X-Y structure out of electrically
connected pads see \fref{fig:2dreadout}. The connections for the formation
of the X-strips are on the one side of the doubly copper clad Kapton, while
the connections for the Y-strips are made at the other side, with the help
of metalized holes on the Y-pads. Each CAST detector has 192 X and 192 Y
strips of $350\,\mu\text{m}$ pitch. The active area therefore is about
$45\,\text{cm}^2$.  The Kapton with the X and Y strips and the readout
lines is glued on a paddle shaped plexiglass piece of the Micromegas
structure, where the readout connectors are also fastened. New improvements
are underway combining an integrated Micromegas and a CMOS micro-pixel
anode plane~\cite{colas:04a,giomataris:06a}.

\subsection{Readout electronics and data acquisition}
The charge on the X or Y strips is read out with the help of four Front End
(FE) electronic cards based on the Gassplex chip~\cite{Santiard:94a}
controlled by a CAEN sequencer (V 551B) with two CRAM (V550) modules in a
VME crate~\cite{Geralis:03a}.

One FE card integrates 96 signals (96 strips) and operates at a maximum
clock speed of 1 MHz. It provides a multilevel output where each level
corresponds to the result of the integration of the signal from a
particular strip. The cards are powered by a 6V power supply (positive and
negative). The Sequencer provides the proper timing signals (Clock, Track
and Hold and Reset) to the FE cards. The CRAM modules integrate and store
the total charge of each channel indicated by the signal provided by the FE
cards until the software reads the data and transfers them to the PC for
permanent storage and analysis. The signal for triggering the Micromegas
device is obtained through the micromesh signal. The output of the
preamplifier is subsequently shaped and amplified to produce the
appropriate trigger signal. Because of the low rates involved (1 Hz) the
zero suppression and pedestal subtraction capabilities of the CAEN modules
are not used and all strip data are recorded.

\begin{figure}
  \begin{minipage}{0.49\textwidth}
    \centerline{\includegraphics[width=0.9\textwidth]{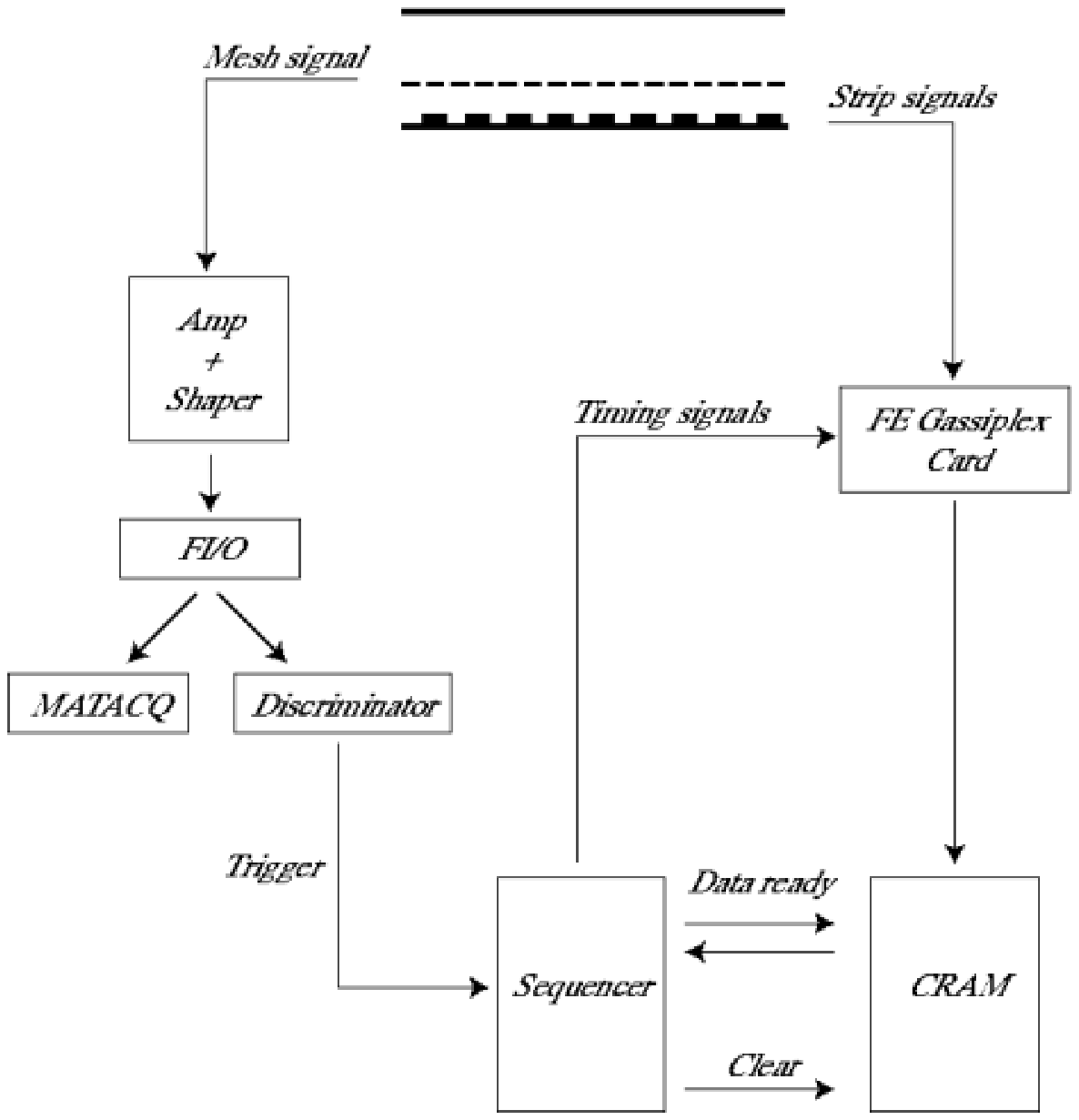}}
    \label{fig:daq}
  \end{minipage}
  \hfill
  \begin{minipage}{0.49\textwidth}
    \centerline{\includegraphics[width=1\textwidth]{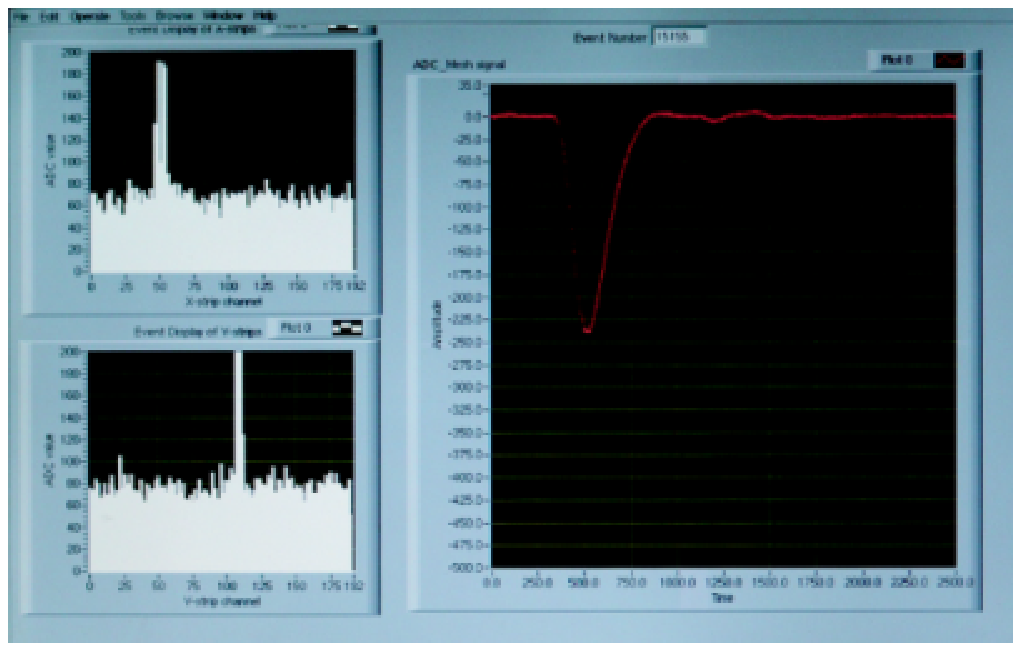}}
    \label{fig:pulsematacq}
  \end{minipage}
  \caption{Left: Trigger and readout logic. Right: Event display showing
    the distribution of X and Y strip charges and the MATACQ pulse.}
\end{figure}

The features of this Micromegas detector also include the recording of the
mesh pulse via a high rate sampling VME Digitizing Board, the MATACQ
Board~\cite{breton:05a}. This board, based on the MATACQ IC, can code 4
analog channels of bandwidth up to 300 MHz over 12 bits dynamic range and a
sampling frequency reaching up to 2 GHz and over 2520 usable points. One of
these channels is used to record the time structure of the mesh pulse.
Signal events have a characteristic mesh pulse that will be used in order
to reject events with unexpected shapes as background events.
\fref{fig:daq} shows a schematic of the Micromegas trigger and readout.
The data acquisition and monitoring system is based on the LabView software
package, of National Instruments, running on a PC with either the Linux
RedHat 7.3.1 (CERN release) or the Windows 2000 operating system. A dual
boot PC is used to connect to the VME Controller and run the data
acquisition software. The connection is performed via a PCI-MXI2 card
sitting on the PCI bus of the PC, a VME-MXI2 controller card sitting on the
VME and a 20~m long MXI2 cable connecting these two cards. The DAQ system
runs on Linux since it provides the facilities of the CASTOR automatic data
archiving system and the xntp software for the synchronization of the PC
clock to the GPS universal time. The online software is controlled by
LabView virtual modules that initialize the run (allowing parameters to be
changed) and monitor its status. An event display is used to view the strip
charges and the mesh signal recorded by the MATACQ board (see
\fref{fig:pulsematacq}). An online analysis is performed in order to give
out plots that are used to monitor the detector performance.
\begin{figure}
\end{figure}

\subsection{Calibrator}
\begin{figure}
  \centerline{\includegraphics[width=7cm]{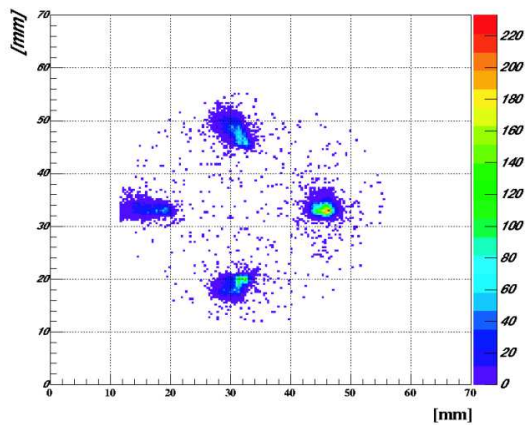}}
  \caption{A two dimensional plot showing the x-y strip image of the four holes
    where the 5.9 keV X-rays go through during a calibration run.}
  \label{fig:calibrationspots}
\end{figure}

The calibration of the detector is done by shining a $^{55}$Fe source daily
at the back of the detector. An automatic mechanism, controlled by the
acquisition is used; the $^{55}$Fe source is moved in front of four blind
holes drilled in the Plexiglass paddle piece to allow the passage of the
5.9 keV X-rays in the chamber (see \fref{fig:calibrationspots}). Once the
calibration run is finished the source is parked inside a shielding.

\section{Detector performance}
\subsection{Characterization}
\begin{figure}
  \begin{minipage}{0.49\textwidth}
    \centerline{\includegraphics[width=0.9\textwidth]{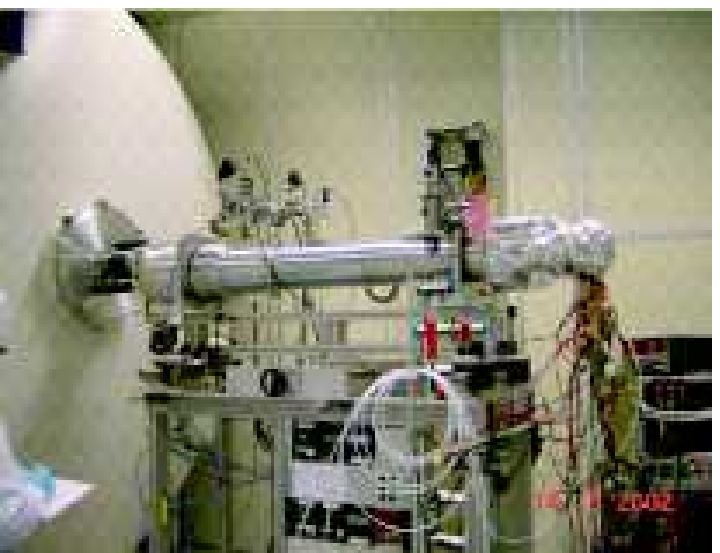}}
  \end{minipage}
  \hfill
  \begin{minipage}{0.49\textwidth}
    \centerline{\includegraphics[width=0.9\textwidth]{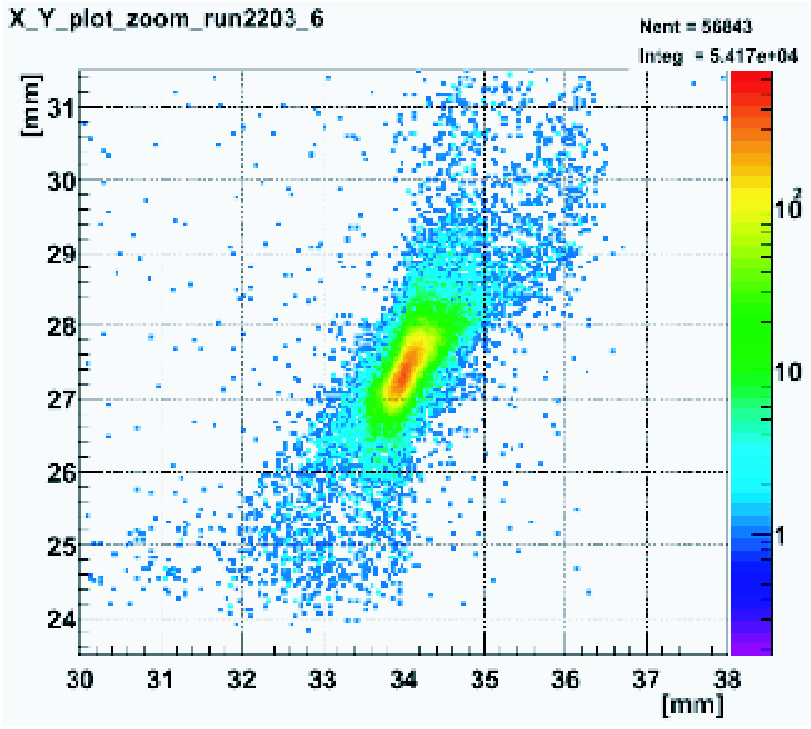}}
  \end{minipage}
  \caption{Left: Photo of the Micromegas detector mounted in the focal
    plane of the X-ray telescope at the PANTER facility. Right: Transverse
    position of the 4.5~keV focused photon beam at PANTER.}
  \label{fig:panterbeam}
\end{figure}

\begin{figure}
  \centerline{\includegraphics[width=8cm]{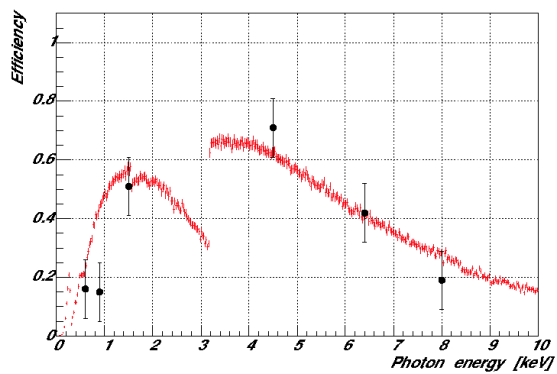}}
  \caption{Simulated detector efficiency as a function of energy with the
    experimental points measured at PANTER.}
  \label{fig:panterefficiency}
\end{figure}

To characterize the detector a test was done at the PANTER X-ray facility
of the Max-Planck-Institut f\"ur extraterrestrische Physik (MPE) in Munich
\cite{freyberg:06a}. A detector was mounted at the X-ray focusing telescope
(now part of the CAST experiment) and tested with photon beams of varying
energy. The detector, at the time, had a buffer space between the vacuum
window and the detector drift electrode filled with Helium gas at
atmospheric pressure. The buffer of Helium gas was used in order to couple
the gaseous Micromegas volume at atmospheric pressure to the vacuum
environment of the X-ray telescope (and of the CAST magnet bore) before the
solution of the differential pumping was adopted. The drift space was 18~mm
wide and the amplification gap was 50~$\mu$m. The X-Y position
determination capability was for the first time shown and the remarkable
agreement with the beam shape expected from the focusing properties of the
X-ray telescope exhibited~\cite{andriamonje:04f}.
Figure~\ref{fig:panterbeam} shows a photo of the experimental set up as
well as the logarithmic intensity plot of the X-Y position of 4.5~keV
photons at the focus. The mm size core of the beam is clearly visible.
The efficiency of the detector was simulated using the GEANT4 package
\cite{geant4}. The dimensions of the detector, the materials of the windows
(drift and helium buffer), the gas mixture as well as the beam spot were
taken into account. In \fref{fig:panterefficiency} the simulated efficiency
with the experimental measured points is shown. The agreement observed
allow us to use this simulation with slightly different parameters (drift
space or window thickness) for the detectors that were used in the data
taking periods.

\subsection{The 2003 detector}
\begin{figure}
  \centerline{\includegraphics[width=7cm]{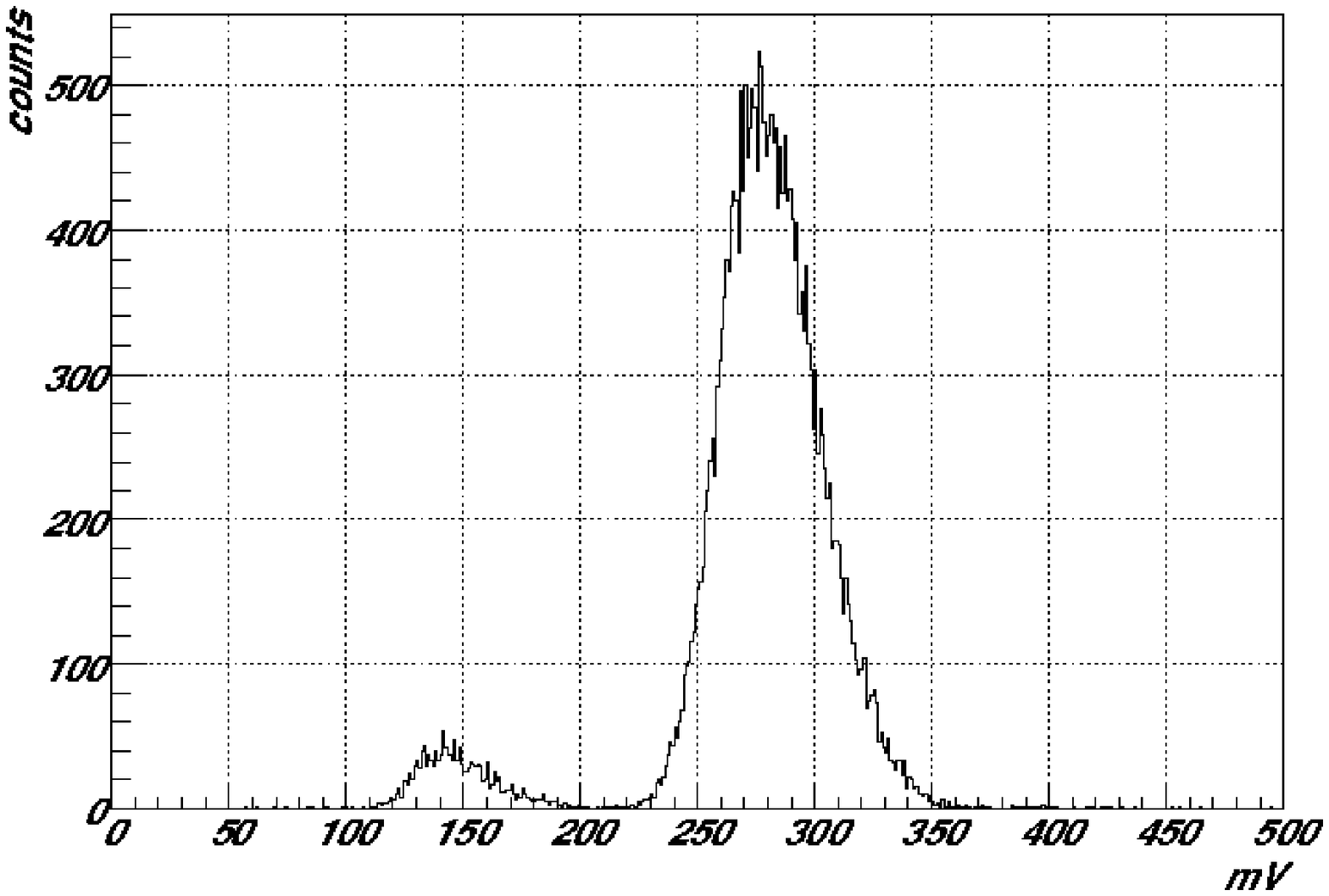}
    \includegraphics[width=7cm]{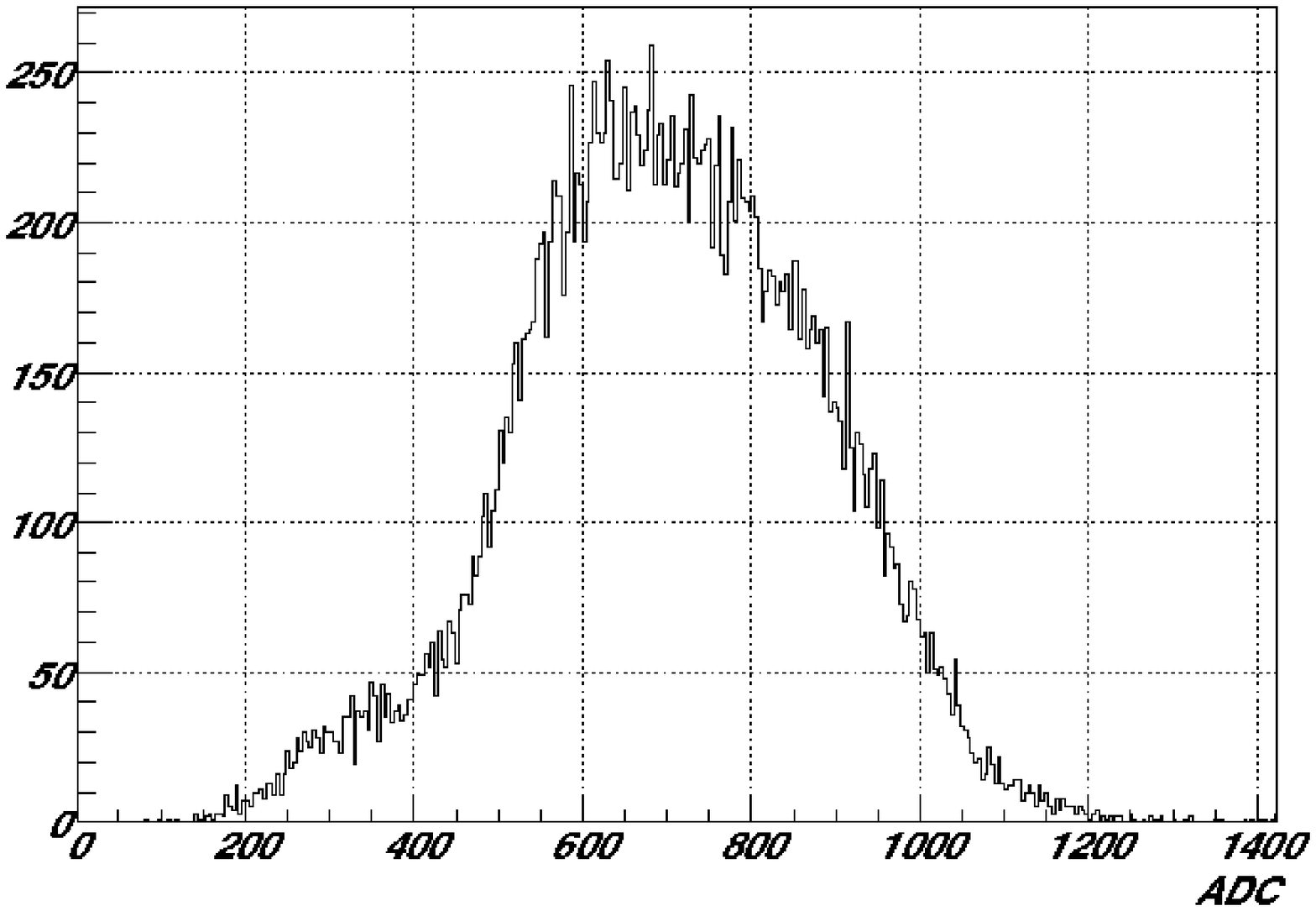}}
  \caption{Energy spectrum for a calibration run using a $^{55}$Fe source
    obtained with the mesh signal read by the MATACQ card (left) and with
    the strips (right).The energy resolution obtained with the strips is
    degraded due to residual crosstalk between the strips.}
  \label{fig:cal2003}
\end{figure}

The Micromegas detector used for the 2003 data taking was designed with a
25~mm drift space and $50\,\mu{\rm m}$ amplification gap is formed by the
help of kapton pillars on the micromesh plane. The detector accumulated
data from May to mid-November without incident. For the last three months
of data taking, the MATACQ card was installed allowing the recording of the
pulse structure of the mesh pulse. An example of a calibration run is given
in \fref{fig:cal2003} where an energy resolution of 16\% (FWHM) is obtained
at 5.9~keV.  The energy resolution obtained with the strips is about 30\%.
This degraded performance was due to some crosstalk between the strips
caused by residual copper left on the kapton pillars of the micromesh which
when in contact with the copper strips of the readout plane gave rise to
this crosstalk.

\begin{figure}
  \centerline{\includegraphics[width=7cm]{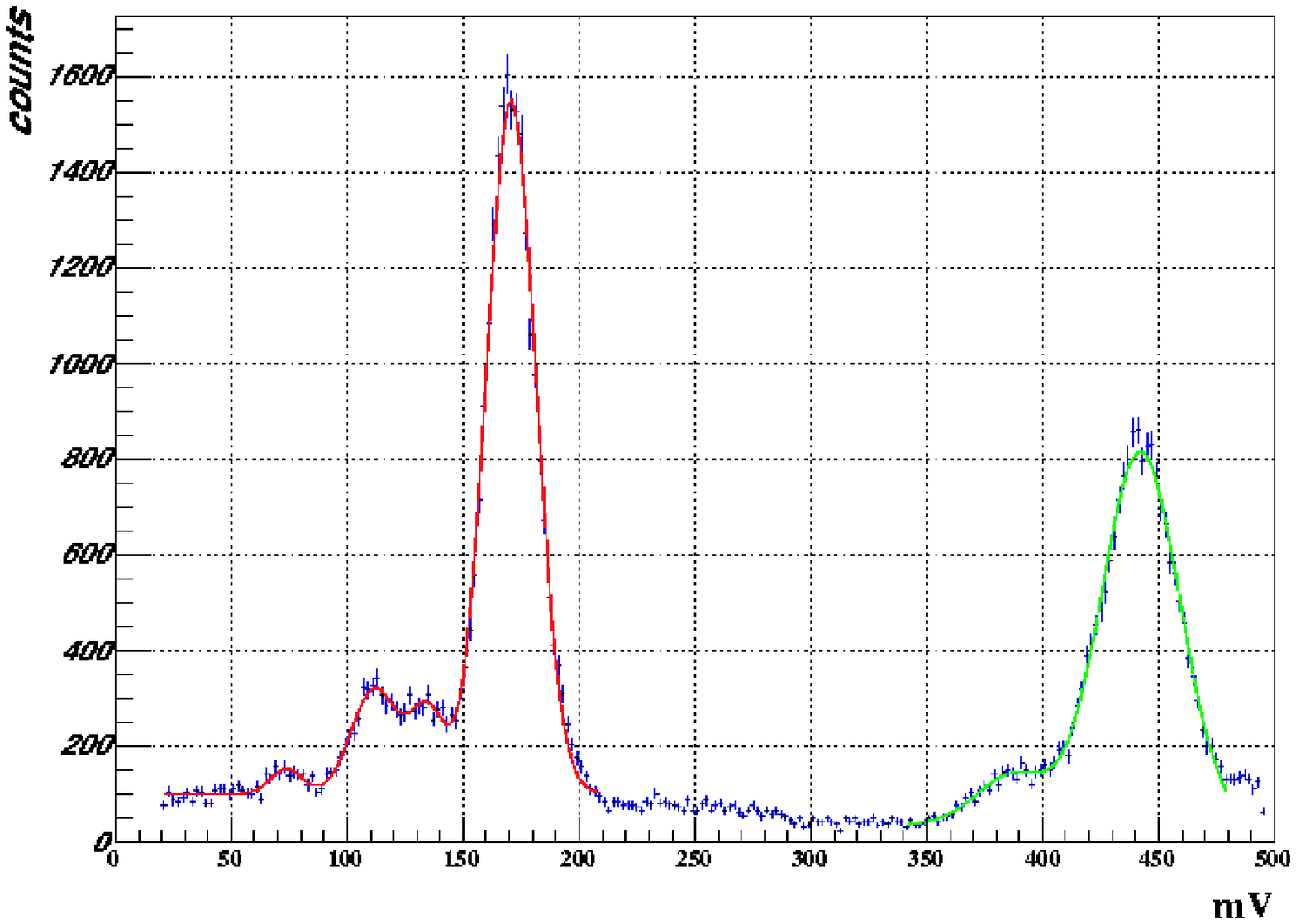}
    \includegraphics[width=7cm]{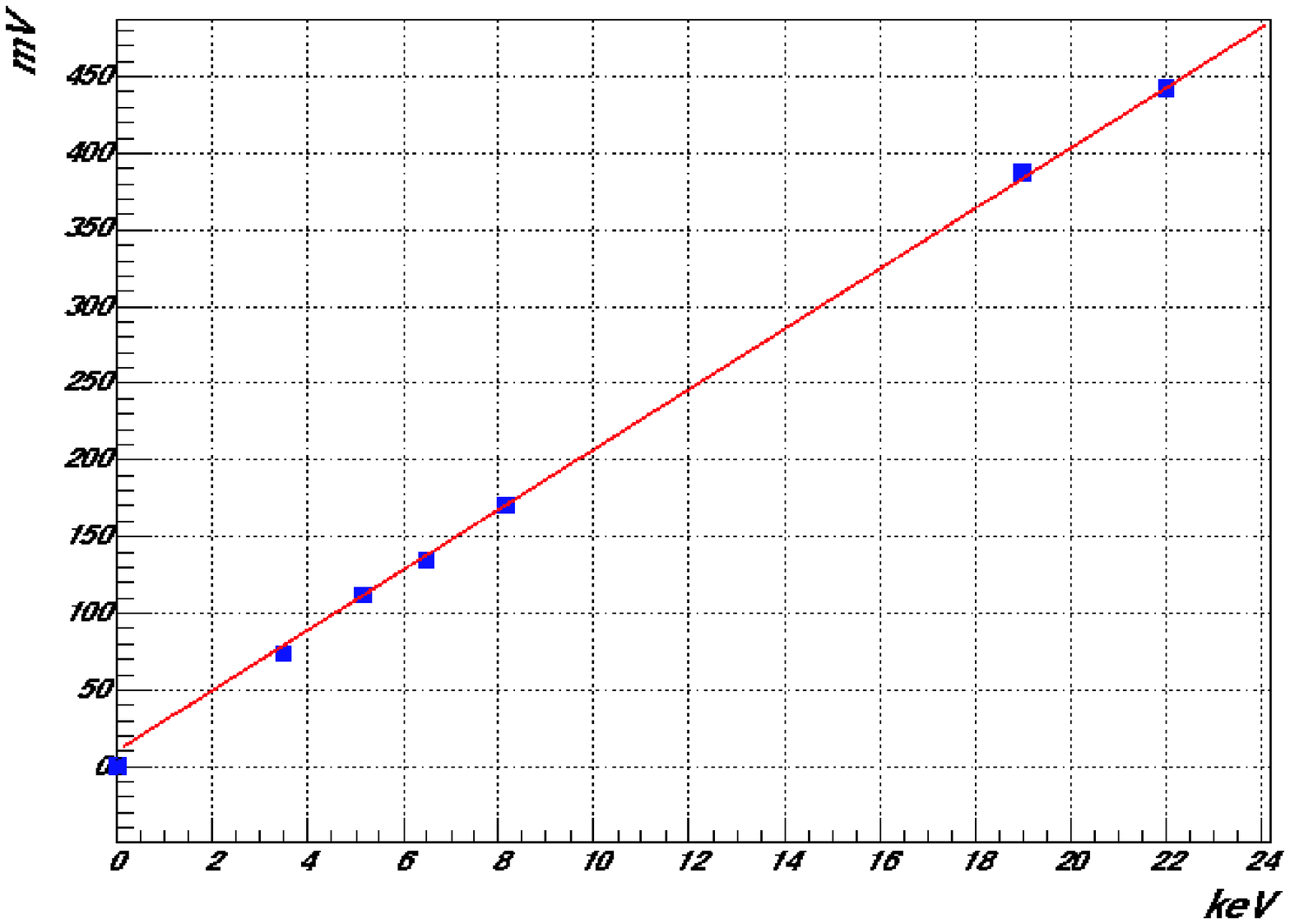}}
  \caption{On the left energy spectra obtained with a Cd Source.
    Peaks at 8~keV due to the fluorescence of the Copper and the Cd 22~keV
    peak can be seen. On the right, the linearity plot showing the pulse
    height as a function of energy.}
  \label{fig:cd}
\end{figure}

The detector's linearity was verified by using a $^{109}$Cd source which
produced fluorescence of the detector's material at different energies.
\Fref{fig:cd} shows the energy spectra as well as the linearity. The system
was extremely stable: the time characteristics and energy response of the
mesh pulses showed less than a 2\% variation during the entire period.

\begin{figure}
  \centerline{\includegraphics[width=8cm]{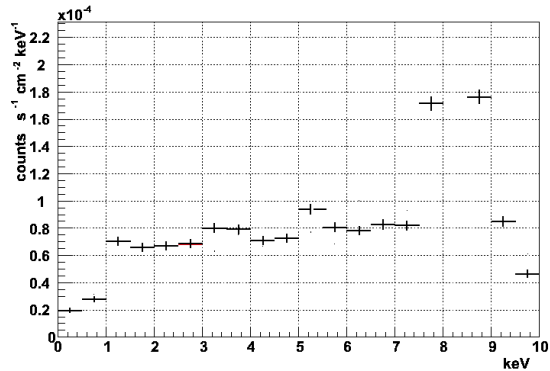}}
  \caption{Background spectra after the filtering cuts for the 2003 data.}
  \label{fig:bkg2003}
\end{figure}

The Micromegas detector records tracking data at sunrise, and during the
rest of the day background data is taken. The detector is calibrated daily.
Signal events (photons with energy of $1$-$8\,\text{keV}$) have a well
defined signature giving a typical cluster in the read out strips and a
typical pulse in the micromesh. Background events, coming from cosmic rays
and natural radioactivity, give a bigger cluster in the strips, and the
pulse shape in the micromesh is very different, favouring an efficient
rejection based on the micromesh pulse shape and on the cluster topology.
The offline analysis was based on sequential cuts, mainly on the micromesh
pulse observables and less on the clustering (due to the strip crosstalk).
\Fref{fig:bkg2003} shows the energy spectra for background events after the
sequential cuts where the average background rate is
$1.4\times10^{-4}\,{\rm counts}\,{\rm cm}^{-2}\,{\rm s}^{-1}\,{\rm
  keV}^{-1}$ region. The background is composed of events coming from
cosmic rays, natural radioactivity and fluorescence from materials present
in the detector. The most visible peak is at 8~keV due to the Copper
present in the anode plane as well as in the mesh cathode. The efficiency
is defined as the ratio of the number of events that pass sequential cuts
over the number of initial reconstructed events before cuts. This
efficiency was calculated using the daily calibration runs giving 80\% and
95\% for 3~keV and 5.9~keV respectively.

\subsection{The 2004 detector}
\begin{figure}
  \centerline{\includegraphics[width=7cm]{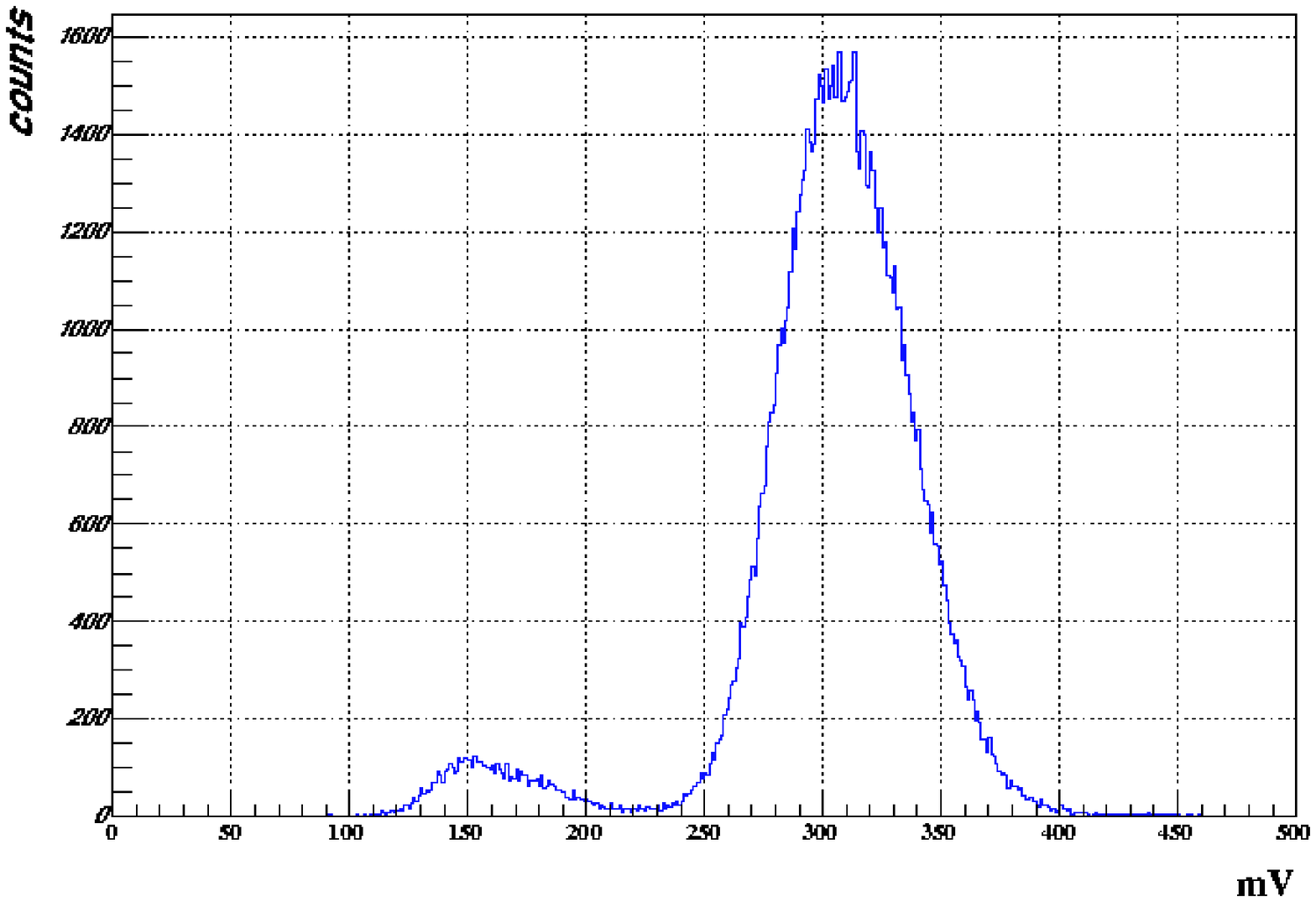}
    \includegraphics[width=7cm]{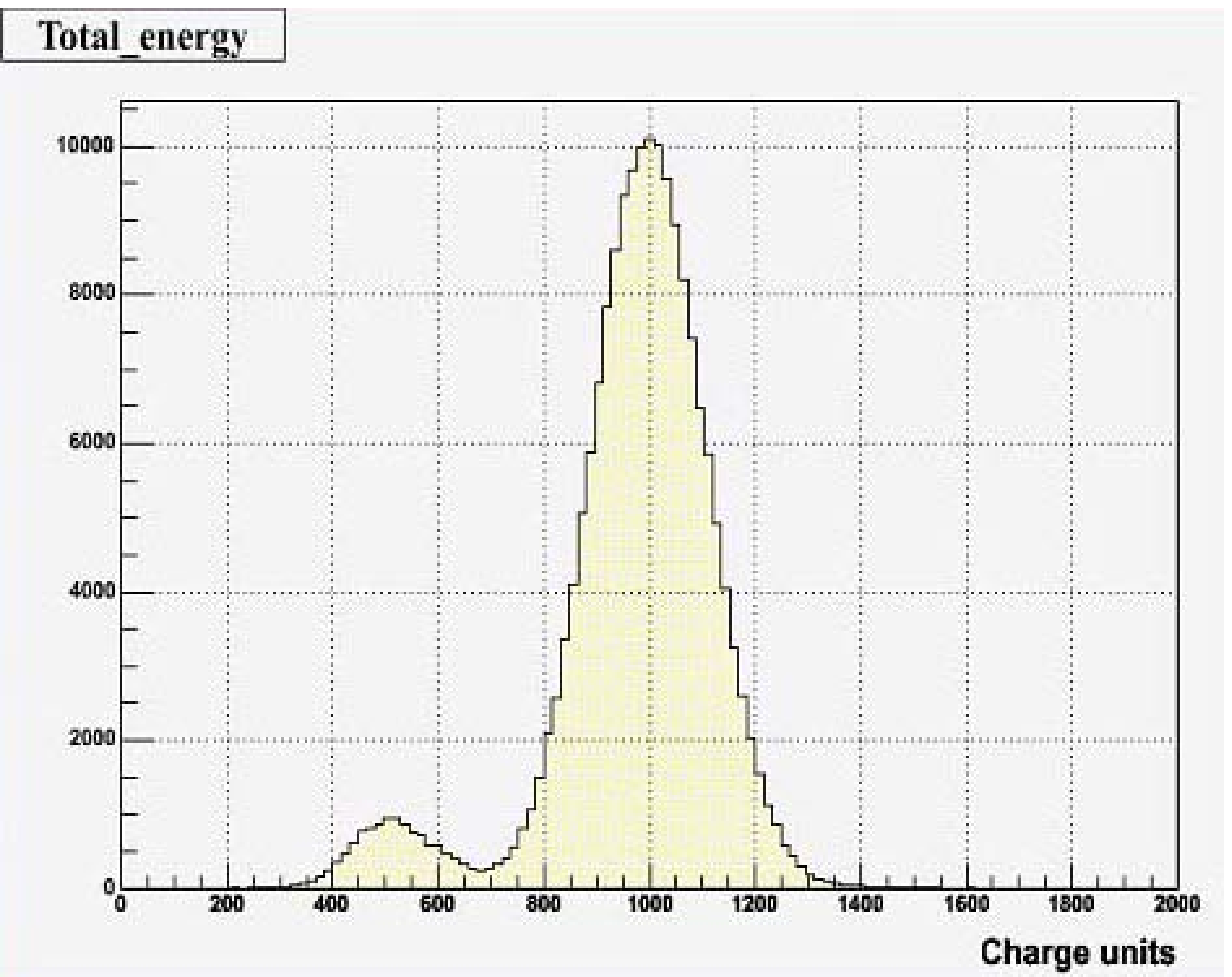}}
  \caption{Energy spectrum for a calibration run using a $^{55}$Fe source
    obtained with the mesh signal read by the MATACQ card (left) and with
    the strips (right). The energy resolution is less than 20\% (FWMH)
    using both signals. }
  \label{fig:cal2004}
\end{figure}

The experience acquired during the 2003 run led to the development of the
V4 model with $30\,$mm conversion gap and $100\,\mu$m amplification gap,
which was designed to eliminate the "crosstalk" effects present at the
previous model and to improve the quality of the strips. Both goals were
achieved and moreover a faster MATACQ board was installed, reducing the
detector's dead time to 14 msec (less than 1.5\% of the net data rate)
while the energy resolution was 19\% FWHM at 5.9 keV. The spectra obtained
with the mesh signal recorded by the MATACQ card (left) and with the strips
(right) are shown in \fref{fig:cal2004}. The energy resolution obtained
with the mesh signal or with the strips is equally good for this detector
due to the reduction of the strips crosstalk.

\begin{figure}
  \centerline{\includegraphics[width=7cm]{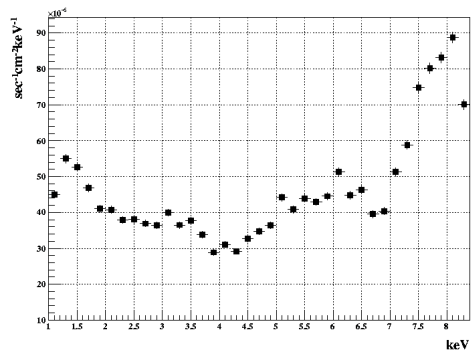}}
  \caption{Background spectra after applying the Fisher discriminant
    method for analysis of the 2004 data
    \citeaffixed{andriamonje:07a}{see}.}
  \label{fig:bkg2004}
\end{figure}

The very accurate strip data allowed us to improve the offline analysis
dramatically by combining the information from the spatial distribution of
the charge collected during an event with the time structure of the mesh
pulses. More specifically, six observables (pulse risetime, pulse width,
pulse height vs pulse integral correlation, X and Y strip multiplicity
balance, X and Y strip charge balance, pulse height vs total strip charge
correlation) were used in a modified Fisher discriminant method to
distinguish more efficiently the proper X-ray events from other signals.
\Fref{fig:bkg2004} shows the resulting background rejection to be at the
level of $4.8\times10^{-5}\,{\rm counts}\,{\rm cm}^{-2}\,{\rm s}^{-1}\,{\rm
  keV}^{-1}$ region with 94\% uniform software efficiency. The system's
stability is demonstrated through the mesh pulses' time structure (0.5\%
variation of risetime and width during the six months of the run) and the
moderate gain variation (10\% on a weekly base) which was corrected with
daily calibration.

\section{Conclusion}
A Micromegas detector with novel features, such as the X-Y strip readout
and the low background materials, was designed and constructed for the
detection of $1$--$10\,{\rm keV}$ X-ray photons for the solar axion search
experiment CAST at CERN. The excellent stability, linearity, position
determination capability, low threshold and good energy resolution are
shown. The analysis of the events permits the rejection of a large fraction
of cosmic ray related background using the observed properties of genuine
photon events such as the rise time and width of the micromesh signal, the
cluster size and the X-Y energy balance. The best background rejection
obtained has been shown to be at the level of $5\times 10^{-5}\,{\rm
  counts}\,{\rm cm}^{-2}\,{\rm s}^{-1}\,{\rm keV}^{-1}$ with an efficiency
of 92\%.  With an appropriate shielding the rejection factor should easily
be improved. This Micromegas design has produced a powerful device for the
detection of X-rays from axions in the energy range of 1-10 keV. The
achieved background rejection opens up the use of the Micromegas detector
for other rare event searches.

\ack

This work has been performed in the CAST collaboration. We thank our
colleagues for their support. Furthermore, the authors acknowledge the
helpful discussions within the network on direct dark matter detection of
the ILIAS integrating activity (Contract number: RII3-CT-2003-506222).

\section*{References}
\bibliography{mnemonic,xmm,cast,conferences,detback,detector,general}
\bibliographystyle{jphysicsB}

\end{document}